\documentstyle[graphicx,12pt,aaspp4]{article}

\def\lsim{\raise0.3ex\hbox{$<$}\kern-0.75em{\lower0.65ex\hbox{$\sim$}}}
\def\gsim{\raise0.3ex\hbox{$>$}\kern-0.75em{\lower0.65ex\hbox{$\sim$}}}

\begin{document}

\title{RING FORMATION IN MAGNETICALLY SUBCRITICAL CLOUDS 
	AND MULTIPLE STAR FORMATION} 

\author{Zhi-Yun Li}
\affil{Department of Astronomy, University of Virginia, P.O. Box 3818}
\affil{Charlottesville, VA 22903; zl4h@virginia.edu}

\begin{abstract}

We study numerically the ambipolar diffusion-driven evolution of 
non-rotating, magnetically subcritical, disk-like molecular clouds, 
assuming axisymmetry. Previous similar studies have concentrated 
on the formation of single magnetically supercritical cores at the 
cloud center, which collapse to form isolated stars. We show that, 
for a cloud with many Jeans masses and a relatively flat mass 
distribution near the center, a magnetically supercritical ring
is produced instead. The supercritical ring contains a mass well 
above the Jeans limit. It is expected to break up, through both 
gravitational and possibly magnetic interchange instabilities, 
into a number of supercritical dense cores, whose dynamic collapse 
may give rise to a burst of star formation. Non-axisymmetric
calculations are needed to follow in detail the expected ring 
fragmentation into multiple cores and the subsequent core
evolution. Implications of our results on multiple star formation 
in general and the northwestern cluster of protostars in the 
Serpens molecular cloud core in particular are discussed.
\end{abstract}

\clearpage

\section{INTRODUCTION}

At the heart of multiple star formation lies cloud fragmentation. The 
role of magnetic fields in the cloud fragmentation is not well 
explored (Boss 2001),
even though present day molecular clouds are thought to be strongly 
magnetized. Reliable Zeeman measurements of the magnetic field strength 
of molecular clouds made to date, as compiled by Crutcher (1999), suggest 
that, after geometric corrections (Shu et al. 1999), the clouds are 
remarkably close to being magnetically critical. These measurements 
reinforce the oft-expressed view that magnetic fields in molecular clouds 
are strong enough to be dynamically important. 

Dynamically important magnetic fields can change the characteristics of 
cloud fragmentation fundamentally. The magnetic and gravitational forces 
are often comparable in magnitude but opposite in direction. The near 
cancellation of forces (Shu \& Li 1997) can in principle keep the clouds 
in a magnetically levitated state over many dynamic times, allowing  
more time for over dense substructures to develop and fragment. 
To isolate the effects of magnetic fields on cloud
fragmentation from those of turbulence (which have been the subject 
of intensive numerical simulations, as reviewed by Vazquez-Semadeni et al.
2000), we will concentrate on the relatively quiescent regions 
of molecular clouds. Such regions are 
capable of producing binaries, multiple stellar systems, as well as 
groups or small clusters - all of which we broadly term ``multiple stars'' 
- but not rich clusters of hundreds to thousands of stars, which tend 
to form in more turbulent regions (Myers 1999). Our goal is to extend 
the reasonably successful scenario of single, isolated star formation, 
based on ambipolar diffusion (Shu, Adams \& Lizano 1987; Mouschovias
\& Ciolek 1999; see, however, Nakano 1998 for a different opinion), to 
the formation of multiple stars. In a longer term, we hope to apply 
the insight gained on the magnetically controlled fragmentation and 
multiple star formation to the more difficult problem of rich cluster 
formation, where turbulence plays a more dominant role (e.g., Klessen, 
Burkert \& Bate 1998). 

In the absence of a strong turbulence, a cloud with many Jeans masses 
supported mainly by an ordered magnetic field tends to settle along 
field lines 
into a disk-like configuration (Mouschovias 1976; Tomisaka, Ikeuchi \& 
Nakamura 1988). The flattened geometry promotes cloud fragmentation 
which, under strict axisymmetry, manifests itself in the formation 
of {\it gravitationally unstable rings}, even in the absence of 
any rotation. This was demonstrated explicitly by Bastien (1983) for 
non-magnetic clouds. Physically, the ring formation results from the 
fact that it would take less time for 
a Jeans mass of material located more than one Jeans length from the 
cloud center to collapse freely onto itself than to reach the center. 
The reason is simply that free fall time is inversely proportional to 
the square root of (average) density.  For a more or less uniform 
mass distribution, the local density at the off-center
location is higher than the average density interior to it, since 
most of the interior is empty for a flattened structure (Bonnell 1999). 
In this paper, we will demonstrate that strong magnetic fields do not 
fundamentally alter the natural tendency of {\it flattened} multi-Jeans 
mass clouds to form rings in the axisymmetric geometry, even though 
the fields provide the bulk of cloud support, flatten the cloud 
material (thus making ring formation possible 
in the first place), and control the pace of cloud evolution 
(through ambipolar diffusion). This behavior is consistent 
with the linear analysis of Langer (1978; see also Pudritz 1990), 
who considered the Jeans instability in an infinite, lightly ionized, 
isothermal medium with a uniform density and magnetic field. 
Employing the standard 
``Jeans swindle'', Langer was able to show that the presence of a 
magnetic field does not change the critical wavelength (and thus 
mass) above which the instability occurs, although the growth 
rate can be strongly modified: 
for typical dark cloud conditions the instability grows on an ambipolar 
diffusion time scale, which is roughly an order of magnitude longer
than the dynamic time scale. One expects the same magnetically-retarded 
Jeans instability to operate in the magnetically supported, 
disk-like clouds of finite extent as well. Indeed, our calculations 
of ring formation in such clouds can be viewed as following the 
nonlinear developments of the instability under the restriction of 
axisymmetry. 

Dense rings sometimes appear in numerical simulations of axisymmetric  
collapse of {\it non-magnetic, rotating} clouds (e.g., Black \& 
Bodenheimer 1976). They are shown to be strongly susceptible to 
dynamical non-axisymmetric fragmentation into small pieces (Norman 
\& Wilson 1980). These rotating rings were 
widely discussed in connection with the production of multiple stellar 
systems in late 1970s and early 1980s, although their formation depends 
sensitively on the numerical treatment of angular momentum 
transport (Norman, Wilson \& Barton 1980). The ring formation to be 
discussed in this paper does not depend on rotation. It represents
a first step towards a theory of multiple star formation in a 
strongly magnetized cloud. 

The rest of the paper is organized as follows. The mathematical 
formulation of the problem of ring formation in a strongly 
magnetized cloud is given in \S~2, and numerical examples are 
presented in \S~3. We discuss in \S~4 ring fragmentation and its 
implications on multiple star formation in general and the 
northwestern cluster of the Serpens molecular cloud core in 
particular. 

\section{FORMULATION OF THE PROBLEM}

\subsection{Governing Equations}
\medskip

We adopt the standard thin-disk approximation (e.g., Nakamura, Hanawa 
\& Nakano 1995) 
and cast the MHD equations that govern the evolution of magnetized 
clouds in a vertically 
integrated form. Mass conservation of the disk material yields
\begin{equation}
{\partial \Sigma\over \partial t} +{1\over r}{\partial\over\partial r}
(r\Sigma V) =0,
\label{e1}
\end{equation}
where $\Sigma$, $t$, $r$, and $V$ are, respectively, the (mass) column 
density, time, cylindrical radius, and the radial component of 
disk velocity. Axisymmetry and a cylindrical coordinate system 
$(r,\phi,z)$ are adopted throughout the paper.  

We assume that the disk is isothermal with an effective sound speed
of $a$ and is threaded by an ordered magnetic field with a 
(cylindrically) radial component $B_r$ and a vertical 
component $B_z$. The vertically integrated momentum equation in the 
radial direction then becomes 
\begin{equation}
{\partial (\Sigma V)\over \partial t}+{1\over r}{\partial\over\partial r}
(r\Sigma V^2)=\Sigma g_r + {B_r B_z\over 2\pi} - {\partial (\Sigma a^2)
\over \partial r} - H {\partial \over \partial r}\left({B_z^2\over 4\pi}
\right),
\label{e2}
\end{equation}
where $g_r$ is the radial component of gravity and $H$ is the disk half 
thickness. The force terms on the right-hand side of the equation are, 
respectively, gravity, magnetic tension, thermal (and possibly turbulent) 
and magnetic pressure force. We have kept only the leading terms for 
the magnetic tension and pressure force, and have ignored cloud rotation, 
which is dynamically unimportant in general before the formation of 
compact stellar objects (i.e., protostars and their disks; see Basu 
\& Mouschovias 1994). Rotation can easily be included if necessary. 

The evolution of the ordered magnetic field is governed by the magnetic
flux conservation equation
\begin{equation}
{\partial B_z\over \partial t}+{1\over r}{\partial\over\partial r}
(r B_z V_{_B})=0,
\label{e3}
\end{equation}
where $V_{_B}$ is the velocity of magnetic field lines in the cross-field 
direction (Nakano 1984). In a lightly ionized medium such as molecular 
cloud, the field lines slip relative to the neutral matter at a 
velocity 
\begin{equation}
V_{_B}-V=t_c \left[{B_r B_z\over 2\pi}-H {\partial \over\partial 
r}\left({B_z^2\over 4\pi}\right)\right]{\bigg /} \Sigma, 
\label{e4}
\end{equation}
where $t_c$ is the coupling time between the magnetic field and 
neutral matter. In the simplest case where the coupling is provided
by ions that are well tied to the field lines and the ion density 
$\rho_i$ is related to the cloud density $\rho$ by the simple 
expression $\rho_i=C\rho^{1/2}$, one has 
\begin{equation}
t_c={1.4\over \gamma C\rho^{1/2}},
\label{e5}
\end{equation}
where typically $\gamma C=1.05\times 10^{-2}$cm$^{3/2}$g$^{-1/2}$
s$^{-1}$ (e.g., Shu 1992) and the factor 1.4 comes from the fact 
that the cross section for ion-helium collision is small compared 
to that of ion-hydrogen collision (Mouschovias \& Morton 1991). 
   
The disk half thickness $H$ in equation (\ref{e2}) and the mass 
density $\rho$ in equation (\ref{e5}) are related through the 
definition
\begin{equation}
\Sigma=2\rho H.
\label{e6}
\end{equation}
To determine these two quantities separately, we assume that the disk
is always in a static equilibrium in the disk-normal direction (Fiedler
\& Mouschovias 1993). Integration of the force balance equation 
vertically yields 
\begin{equation}
\rho={\pi G \Sigma^2\over 2 a^2}\left(1+{B_r^2\over 4\pi^2 G\Sigma^2}
\right) + {P_{\rm e}\over a^2},
\label{e7}
\end{equation}
to the lowest order in $(H/r)$. The two terms in the brackets
represent, respectively, the gravitational compression and 
magnetic squeezing of the disk material. The magnetic squeezing 
term becomes important late in the cloud evolution, when the 
magnetic field configuration becomes highly pinched. The quantity 
$P_{\rm e}$ denotes the ambient pressure that helps confine the 
disk, especially in low column density regions where gravitational 
compression is relatively weak. 

\subsection{Initial Conditions}
\label{initial}

The ``initial'' distributions of mass and magnetic flux of a star-forming
cloud are not well determined either observationally or theoretically. 
For illustrative purposes, we prescribe them in a ``reference'' state,
following Basu \& Mouschovias (1994). We adopt a uniform distribution 
\begin{equation}
B_z^{\rm ref}(r)=B_\infty,
\label{e8}
\end{equation}
everywhere for the magnetic field, with $B_\infty$ denoting the background 
field strength, and a simple prescription 
\begin{equation}
\Sigma^{\rm ref}(r)={\Sigma_0\over \left[1+\left({r/r_0}\right)^n
\right]^{4/n} },
\label{e9}
\end{equation}
for the column density, with $\Sigma_0$ denoting the central value
and $r_0$ a characteristic radius beyond which the column density 
drops off rapidly. We will sometimes refer to $\Sigma_0$ simply as 
``the reference column density'' later. The prescriptions (\ref{e8}) 
and (\ref{e9}) are similar to those used by Basu \& Mouschovias 
(1994), except that we leave the exponent $n$ free to specify. 
The exponent $n$ controls the amount of mass in the central 
``plateau'' region where the mass distribution is more or less 
uniform. It will play a crucial role in ring formation (\S~3). The 
low-column density ``envelope'' outside the radius $r_0$ is 
designed mainly to minimize the effects of cloud boundary on the 
evolution of the central region. For the model clouds to be 
considered in the next section, we will adopt a single cloud 
radius that is twice the characteristic radius $r_0$. The column 
density at the cloud edge will then be less than 1/16 of the 
central value in all cases. A region with such a low column density 
will be firmly controlled by magnetic fields and its evolution will 
be effectively decoupled from that of the central region, where 
dynamic collapse and star formation occur.

The reference model clouds prescribed by equations (\ref{e8}) and 
(\ref{e9}) are not necessarily in mechanical equilibrium, since 
the thermal and magnetic forces are not designed to balance out 
the self-gravity exactly. We let these clouds adjust towards an 
equilibrium configuration under the constraints (a) that the magnetic 
field be frozen in matter, and (b) that the field strength remain
fixed at the cloud edge (so that continuity with the background 
field is preserved, see Basu \& Mouschovias 1994). The first 
constraint guarantees that mass-to-flux ratio is conserved for 
each individual mass element during the adjustment. The second 
constraint, together with the first, implies that the column 
density at the cloud edge remains fixed as well. Numerically, the 
equilibrium configurations are obtained by evolving the reference 
clouds in time according to the governing equations 
({\ref{e1})-(\ref{e3}), setting the velocity of magnetic field lines 
$V_{_B}$ equal to the velocity $V$ of neutral matter. An extra 
damping force proportional to $-V$ is applied to the right hand 
side of the momentum equation ({\ref{e2}) during this adjustment 
phase to gradually bring the reference clouds into a static 
equilibrium. The final equilibrium configurations, in which $V=0$ 
and the self-gravity is balanced exactly by a combination of thermal 
and magnetic forces, serve as the initial configurations for the 
subsequent cloud evolution driven by ambipolar diffusion. For the 
magnetically sub-critical clouds that we are interested in, the 
adjustment is usually rather small, as noted previously by Basu 
\& Mouschovias (1994). 

\subsection{Dimensional Units and Dimensionless Quantities}
\label{units}

The governing equations (\ref{e1})-(\ref{e3}) are to be solved 
numerically. To facilitate the numerical attack, we first 
cast all dimensional quantities in these equations into a 
dimensionless form. We begin by normalizing the column density 
and radius with the central value $\Sigma_0$ and characteristic
radius $r_0$ that appear in the reference column density 
distribution, equation (\ref{e9}), and denote the resultant 
dimensionless quantities by 
\begin{equation}
\sigma\equiv{\Sigma\over \Sigma_0}; \ \ \ \ \xi\equiv{r\over r_0}.
\label{e10}
\end{equation}
For typical values of $\Sigma_0$ and $r_0$, we adopt $10^{-2}$ g cm$^{-2}$ 
(corresponding to a molecular hydrogen number column density of $2.1\times 
10^{21}$cm$^{-2}$, or roughly two magnitudes of mean visual extinction,
e.g., McKee 1989) and 1 pc. These typical values introduce two scaling 
factors $\mu_\Sigma\equiv \Sigma_0/(10^{-2}$ g cm$^{-2}$) and $\mu_r\equiv 
r_0$/(1 pc). The dimensional units for other quantities 
are then obtained from various combinations of $\Sigma_0$ and $r_0$. 
In terms of the scaling factors $\mu_\Sigma$ and $\mu_r$, we have the 
following: 
$M_0=2\pi \Sigma_0 r_0^2=3.0\times 10^2 (\mu_\Sigma\mu_r^2)\ {\rm M}_\odot$ 
for mass, $V_0=(GM_0/r_0)^{1/2}=1.1\ (\mu_r\mu_\Sigma)^{1/2}$ km s$^{-1}$ 
for velocity, 
$t_0=r_0/V_0=8.6\times 10^5 (\mu_r/\mu_\Sigma)^{1/2}$ years for 
time, $B_\infty=\Gamma (2\pi G^{1/2}\Sigma_0)
=16\ \Gamma\ \mu_\Sigma$ $\mu$G for field strength, 
$\rho_0=\Sigma_0/r_0=3.2\times 10^{-21} (\mu_\Sigma/\mu_r)$ 
g cm$^{-3}$ for mass density, and $P_0=\pi G\Sigma_0^2/2=
1.0\times 10^{-11}\mu_\Sigma^2$ dyn cm$^{-2}$ for pressure. Note 
that the dimensionless parameter $\Gamma$ is the ratio of the 
background field strength to the critical field strength, $2\pi 
G^{1/2}\Sigma_0$, associated with the reference column density 
$\Sigma_0$. It is a free parameter that characterizes the degree 
of cloud magnetization. Note also that the units $t_0$ and $V_0$ 
are the characteristic free fall time and free fall speed of the 
reference cloud.

With the above defined units, we obtain the following dimensionless 
quantities
$$
m\equiv{M\over M_0};\ \ h\equiv{H\over r_0};\ \ b_r\equiv
{B_r\over B_\infty};\ \ 
b_z\equiv{B_z\over B_\infty}; \ \ \tau\equiv{t\over t_0};
$$
\begin{equation}
v\equiv{V\over V_0}; \ \ v_{_B}\equiv{V_B\over V_0};
\ \ {\hat a}\equiv{a\over V_0}; \ \ {\hat g}_r\equiv{g_r\over V_0^2/r_0}; 
\ \ {\hat \rho}\equiv{\rho\over \rho_0}; 
\ \ p_{\rm e}\equiv{P_{\rm e}\over P_0}.
\label{e11}
\end{equation}
The dimensionless effective sound speed ${\hat a}$, which will play an
important role in ring formation (\S\ref{ring}), can be written in 
terms of the effective cloud temperature $T_{\rm eff}$ as
\begin{equation}
{\hat a}=0.29\ \left({1\over \mu_r\mu_\Sigma}\right)^{1/2} \left({T_{\rm 
eff}\over 30\ {\rm K}}\right)^{1/2},
\label{e12}
\end{equation}
where a helium abundance of $10\%$ by number has been assumed. 
Combining equations (\ref{e7}) and (\ref{e12}), we find a 
molecular hydrogen number density of 
\begin{equation}
n_{_{{\rm H}_2}}^{\rm{ref}}=2.2\times 10^3\ (1+p_e)\ \mu_\Sigma^2
\left({30\ {\rm K}\over T_{\rm eff}}\right)\ \ {\rm cm}^{-3},
\label{e13}
\end{equation}
at the center of the reference clouds, where $\Sigma=\Sigma_0$ and 
$B_r=0$ (by symmetry). In the fiducial case of $\mu_\Sigma=1$, 
$T_{\rm eff}=30$ K and $p_e\ll 1$ (i.e., gravity dominating external 
pressure in compressing the cloud center), the above 
number density is typical of the $^{13}$CO clumps found in many 
molecular clouds. The clumps are often taken as the starting point 
of isolated star formation calculations. 
For cluster forming regions, somewhat higher reference density and
effective temperature may be more appropriate. 

\subsection{Dimensionless Governing Equations and Boundary Conditions}

We rewrite the governing equations of cloud 
evolution into a dimensionless, {\it Lagrangian} form using the 
dimensionless time $\tau$ and mass $m$ as independent variables: 
\begin{equation}
{\partial \xi\over\partial m}={1\over \sigma\xi},
\label{e14}
\end{equation}
\begin{equation}
{\partial \xi\over\partial \tau}=v,
\label{e15}
\end{equation}
\begin{equation}
{\partial v\over \partial \tau}={\hat g}_r+{\Gamma^2 b_r b_z\over
	\sigma} - {\hat a}^2\xi {\partial\sigma\over \partial m}
	-\Gamma^2 h\xi {\partial\over\partial m}
	\left({b_z^2\over 2}\right),
\label{e16}
\end{equation}
\begin{equation}
{\partial\over\partial \tau}\left({b_z\over\sigma}\right)=
-{\partial\over\partial m}[\xi b_z(v_{_B}-v)].
\label{e17}
\end{equation}
The dimensionless drift velocity between magnetic field lines and 
neutral cloud matter that appears in the field diffusion equation
(\ref{e17}) is given by  
\begin{equation}
v_{_B}-v={\Gamma^2 \over \nu_c {\hat \rho}^{1/2}}
	\left[{b_r b_z\over \sigma} 
	-h \xi {\partial\over\partial m}\left({b_z^2\over 2}
	\right)\right],
\label{e18}
\end{equation}
where the magnetic coupling coefficient $\nu_c$ has a value of $11.6$ 
for the simplest case of coupling given by equation (\ref{e5}). 
The dimensionless half thickness and mass density of the disk are 
determined from
\begin{equation}
h={2{\hat a}^2\over \sigma}\left(1+{\Gamma^2 b_r^2+p_{\rm e}\over \sigma^2}
\right)^{-1},
\label{e19}
\end{equation}
and 
\begin{equation}
{\hat \rho}={\sigma^2\over 4{\hat a}^2}\left(1+{\Gamma^2  
b_r^2+p_{\rm e}\over\sigma^2}\right).
\label{e20}
\end{equation}
Together with the auxiliary equations (\ref{e18})-(\ref{e20}), the four 
governing equations (\ref{e14})-(\ref{e17}) completely determine the 
time evolution of four cloud quantities: column density $\sigma$, 
radius $\xi$, radial velocity $v$ and the vertical field strength $b_z$, 
provided that the radial component of gravity ${\hat g}_r$ and the 
radial field strength $b_r$ are determined. These two quantities are 
determined approximately as follows.

It is well known that the radial component of gravity on an infinitely 
thin disk with column density $\sigma(\xi)$ is given by
\begin{equation}
{\hat g}_r=\int_0^\infty \xi^\prime\sigma(\xi^\prime)
M(\xi,\xi^\prime)\ d\xi^\prime,
\label{e21}
\end{equation}
where the integral kernel is obtained from
\begin{equation}
M(\xi,\xi^\prime)=\left({2\over \pi}\right){d\over d\xi}\left[{1\over 
	\xi_{\rm{max}}} K\left({\xi_{\rm{min}}\over\xi_{\rm{max}}}
	\right)\right],
\label{e22}
\end{equation}
with $\xi_{\rm{min}}=$min$(\xi,\xi^\prime)$, $\xi_{\rm{max}}=$max$(\xi,
\xi^\prime)$ and $K$ being the complete elliptic integral of the 
first kind. Following Ciolek \& Mouschovias (1993), we shall use 
equation (\ref{e21}) to approximate the radial component of gravity 
for our thin disk of finite thickness. 

On an infinitely thin disk, the radial field strength $b_r$ can 
be determined from the vertical field strength $b_z$. We adopt the 
formalism of Lubow, Papaloizou \& Pringle (1993), who assumed that 
the magnetic field consists of two parts: a uniform background 
and a potential field due to a toroidal current confined entirely 
to the disk. The first part is prescribed, and the second part is 
determined solely by the (vertically-integrated) toroidal current 
density $J_\phi$, which is linearly proportional to $b_r$ on the 
disk. In particular, $b_z$ on the disk can be expressed as an 
integral over disk radius with an integrand that is linearly 
proportional to $J_\phi$ (Jackson 1975), and thus $b_r$. A simple 
matrix inversion then allows one to calculate $b_r$ in terms of 
$b_z$ (see Lubow et al. 1993 for details). As an approximation, we 
shall use the above formalism to compute the radial field strength
for our thin disk of finite thickness.  

For boundary conditions, we impose at the center of the cloud (where
$m=0$) the conditions of symmetry: 
$\xi=0$, $v=0$, $\partial b_z/\partial m=0$, and $\partial \sigma
/\partial m=0$. At the outer edge, we let the cloud boundary move
freely while holding the vertical component of magnetic field 
$b_z$ and the column density of the disk $\sigma$ fixed at their 
initial values at all times, consistent with the boundary conditions
we imposed in \S~2.2 during the adjustment from the reference state 
towards the equilibrium configuration. These specifications complete 
our discussion of governing equations, initial and boundary 
conditions.

\subsection{Numerical Method}

The numerical methods for Lagrangian hydrodynamics and 
magnetohydrodynamics are well documented in Chapters 4 and 8
of Bowers \& Wilson (1991), respectively. We follow the procedures 
outlined in those chapters closely, except for the treatment of 
magnetic field diffusion equation (\ref{e17}). The field diffusion 
is driven by a combination of magnetic tension and pressure forces, 
corresponding to the two terms on the right hand side of equation 
(\ref{e18}). We use the method of operator splitting to treat the 
two driving terms separately. While the usual finite differencing 
is adequate for the tension term, special treatment is necessary 
for the magnetic pressure term to ensure stability; for the pressure 
term, we employ the so-called ``Gaussian elimination'' technique, 
as described in section 6.4 of Bowers \& Wilson (1991). 

\section{RING FORMATION IN MAGNETICALLY SUBCRITICAL CLOUDS}
\label{ring}

In this section, we numerically integrate the non-dimensional governing 
equations (\ref{e14})-(\ref{e17}), subject to the initial and boundary 
conditions described in \S~2.2 and \S~2.4. Two dimensionless constants 
appear in the governing equations: $\Gamma$, the ratio of the 
background field strength $B_\infty$ to the critical field strength
$2\pi G^{1/2} \Sigma_0$, and ${\hat a}$, the dimensionless effective 
sound speed. The quantity 
$\Gamma$ controls the degree of cloud magnetization, and must be greater 
than unity in order for the clouds to be mainly supported by (ordered)
magnetic fields. Although Nakano (1998) presented theoretical arguments 
against star-forming clumps being magnetically subcritical, the 
situation is less clear observationally: 
Zeeman measurements of the field strength in molecular clouds, as 
compiled by Crutcher (1999), is roughly consistent with the clouds 
being magnetically critical, after likely geometric corrections 
(Shu et al. 1999). Uncertainties involved in estimating the 
mass-to-flux ratio preclude a firmer conclusion. We believe that at 
least some star-forming clumps are magnetically subcritical {\it to 
begin with}\footnote{We note that some Zeeman measurements are for 
{\it evolved} regions where the mass-to-flux ratio may have already 
been increased through ambipolar diffusion to supercritical values 
(McKee 1999).}, and adopt for definitiveness a round value of $\Gamma=2$. 
Clouds with values of $\Gamma$ substantially less than 2 will not be 
disk-like and the thin-disk approximation adopted here may not be 
applicable. Values of $\Gamma$ much larger than 2, on the other hand, 
are difficult to reconcile with the Zeeman measurements. For the 
dimensionless effective sound speed, we will first adopt a round value 
of ${\hat a}=0.3$ (corresponding to an effective temperature of 32 K 
for the canonical choice of the scaling factors $\mu_r=\mu_\Sigma=1$), 
and then consider a smaller value of ${\hat a}=0.2$ for comparison. A
third constant, the dimensionless external pressure
$p_{\rm e}$, appears in auxiliary equations (\ref{e19}) and
(\ref{e20}). We adopt, for simplicity, a value of $p_{\rm e}=0.1$,
which corresponds to a reasonable dimensional value of $10^{-12}
\mu_\Sigma^2$ dyn cm$^{-2}$.  

Three model clouds are considered. They have, respectively, $n=2$, 4 
and 8, where $n$ is the exponent that specifies the column density 
distribution in the reference state.  A quick inspection of the
prescription of the reference column density distribution, equation 
(\ref{e9}), reveals that the $n=8$ cloud has the largest cloud 
mass for the same cloud radius (taken to be twice the characteristic 
radius $r_0$ for all clouds, as mentioned earlier). Its dimensionless 
mass is $0.80$, compared with $0.40$ ($0.66$) for the $n=2$ ($4$) 
cloud. The Jeans mass is, on the other hand, much smaller, at least 
near the cloud center, where 
\begin{equation}
M_J={1.17 a^4\over G^2\Sigma_0},
\label{e240}
\end{equation}
according to Larson (1985; his equation [9]). The dimensionless central 
Jeans mass takes the following simple form
\begin{equation}
m_J=7.35 {\hat a}^4,
\label{e250}
\end{equation}
which has a value of $0.06$ for our standard choice of dimensionless
sound speed ${\hat a}=0.3$. 

The reference clouds are allowed to settle towards an equilibrium 
configuration under the influence of an artificial damping force, 
with magnetic flux frozen-in, as described in \S\ref{initial}. 
After the equilibrium state is 
reached, we reset the time $t$ to zero and turn on ambipolar diffusion. 
The clouds then evolve on the magnetic flux redistribution time scale, 
which is typically an order of magnitude longer than the dynamic 
time scale. In what follows, we will illustrate the main features 
of cloud evolution by displaying various cloud quantities 
at the initial equilibrium time $t=0$ and two representative 
times, $t_1$ and $t_2$, when the maximum column density reaches,
respectively, $10$ and $10^2$ times the reference value $\Sigma_0$. 
Typically, by the time $t_1$,  
the clouds start to become magnetically supercritical, and are
about to collapse dynamically. By the time $t_2$, the collapse
is well into the dynamic phase, with a maximum infall speed 
comparable to, or greater than, the effective sound speed $a$. 

First, we plot the distributions of column density and infall velocity
at the time $t=0$ and $t_1$ in Fig.~1. The latter time corresponds 
to 6.75, 7.87, and 8.63 million years \footnote{We set the scaling 
factor $\mu_r$ and $\mu_\Sigma$ to unity hereafter (including in
the figures) to obtain concrete dimensional quantities from the 
dimensionless numerical solutions. The dependences of dimensional
units on $\mu_r$ and $\mu_\Sigma$ are given in \S\ref{units}. } for 
the $n=2$, $4$, and $8$ cloud, respectively. The column density 
distributions at time $t_1$ illustrate clearly two distinctive modes
of cloud evolution: in the $n=2$ cloud with a relatively peaky
initial mass distribution (see the insert), a dense core has formed 
at the center. The more massive $n=8$ cloud with a flatter central 
mass distribution has produced, on the other hand, a dense ring. The
intermediate $n=4$ cloud is close to the borderline between 
core-forming and ring-forming; a slight increase of the exponent 
$n$, to a value of 4.2 for example, would induce the cloud to form 
a ring instead of a core. 

Both the central cores and dense ring are formed quasi-statically, as 
indicated by the velocity distributions shown in panel (b) of Fig.~1. 
Although the maximum infall speeds are substantial, ranging from 
$\sim 0.05$ to $\sim 0.22$ km s$^{-1}$ in our particular examples, 
they remain below the effective sound speed (and should be even 
smaller compared with the magneto-sonic speed, the relevant signal 
speed in a magnetized cloud). Note that the (subsonic) infall regions
are clearly ``large scale'', spanning a good fraction of a parsec, 
even during this relatively early ``starless'' phase of evolution 
(see also Li 1999 and Ciolek \& Basu 2000). 
Such an extended infall motion may have been detected in the starless 
core L1544 (Taffala et al. 1998; Williams et al. 1999). For starless 
cores formed in magnetically subcritical clouds, Ciolek \& Basu 
(2000) showed that the infall speed is sensitive to the background 
field strength (i.e., the value of $\Gamma$ in our notation). Panel 
(b) demonstrates that the speed depends rather strongly on the initial 
mass distribution as well. 

Second, we plot, in Fig.~2, the distributions of mass-to-flux ratio
and the cloud shapes at the time $t=0$ and $t_2$. The latter time
corresponds to 6.95, 8.10, and 8.94 million years for the $n=2$, $4$, 
and $8$ cloud, respectively. By the time $t_2$, a substantial amount
of mass (14, 55, and 90 $M_\odot$, respectively) has become magnetically 
supercritical, inside either the central core or dense ring,
according to panel (a). Note that the mass-to-flux ratio of the
$n=8$ cloud peaks inside the ring, and that the central region 
interior to the ring remains magnetically subcritical. The relatively 
strong central field should cushion the contraction of the ring 
towards the origin as a whole, with potentially observable signatures. 

The cloud shapes, as outlined by the half thickness $H$ defined in 
equation (6) and shown in panels (b)-(d) of Fig.~2, change 
dramatically as the clouds evolve. The minimum thickness occurs at 
the column density maximum inside either the central cores or dense
ring, where self-gravity squeezes on the disk material the 
hardest. The lower-column density ``envelope'' is compressed, in 
contrast, mainly by external pressure. Together, the self-gravity 
and external pressure keep the cloud material flattened 
at all times, justifying the thin-disk approximation adopted. 
Also shown in panels (b)-(d) are magnetic field lines at time 
$t_2$. It is clear that mass accumulation in the core ($n=2$ and 4) 
and the ring ($n=8$) has also led to an accumulation of magnetic 
flux in these over dense regions, creating a pinched magnetic 
configuration. Indeed, the mass-to-flux ratio at time $t_2$ 
remains less than twice the critical value everywhere (see panel 
[a]), even though the column density has 
increased by a factor of nearly $10^2$ from its initial value, and 
the volume density by an even larger factor, close to $10^4$. The 
near critical mass-to-flux ratio and associated strong magnetic 
field are characteristic of dense cores (see, e.g., Lizano \& Shu
1989 and Basu \& Mouschovias 1994) and rings formed out of 
magnetically subcritical clouds, in contrast with those formed 
in other (weakly- or non-magnetic) scenarios.

The dichotomy of core and ring formation is illustrated most vividly 
by Fig.~3, where the column density distributions of the core-forming
$n=2$ and ring-forming $n=8$ cloud are represented graphically at the 
time $t=0$, $t_1$ and $t_2$. The dense, opaque central core of the
$n=2$ cloud is expected to collapse quickly, on a (very short, local) 
dynamic time scale, to produce a single, isolated star. What happens 
to the ring? In \S\ref{fragmentation} below, we will argue that the 
dense, self-gravitating ring is likely to fragment into smaller pieces, 
which collapse to form more than one star. Therefore, the dichotomy 
of core and ring formation should be indicative of two modes of star 
formation in a strongly magnetized cloud: a single, isolated mode and 
a multiple, clustered mode, depending on the mass of the cloud and 
its distribution.  

So far, we have concentrated on how mass distribution, as specified 
by the exponent $n$ in equation (\ref{e9}), affects core and ring 
formation. We now wish to demonstrate that the effective sound 
speed, $a$, has a profound effect as well. For this purpose, we 
repeat the evolution of the same three clouds shown in Figs.~1-3 
with a smaller value of dimensionless sound speed ${\hat a}=0.2$ 
(instead of 0.3). We find 
that, whereas the $n=2$ ($n=8$) cloud forms a central core (dense 
ring) as before, the borderline $n=4$ cloud is induced by the lower 
sound speed to collapse off-center into a ring instead of a central 
core. This difference is consistent with the trend found by Bastien 
(1983) for flattened, {\it non-magnetic} clouds: decreasing the sound 
speed (and thus the Jeans mass) makes ring formation easier. Indeed, 
if we were to remove all of the magnetic fields from our magnetically 
supported clouds suddenly (assuming ${\hat a}=0.2$), they would 
collapse promptly, with the $n=2$ cloud forming a dense core and the 
$n=4$ and $8$ cloud each forming a dense ring, just as their magnetized 
counterparts. The strong magnetic fields lengthen the ring formation 
time (by roughly an order of magnitude), but do not appear to 
suppress the ring formation tendency of a multi-Jeans mass cloud
with a relatively flat mass distribution. Indeed, it is the magnetic 
fields that are responsible for the cloud flattening and thus the ring 
formation in the first place. We have explored other forms of mass 
and magnetic flux distribution as well as other values of the 
dimensionless sound speed ${\hat a}$, and come to the same general 
conclusion. 

Finally, we note that Mouschovias and collaborators have studied the 
ambipolar diffusion-driven evolution of disk-like magnetic clouds 
extensively (e.g., Ciolek \& 
Mouschovias 1994; Basu \& Mouschovias 1994, 1995), although none of 
their models produced a dense ring. The lack of ring formation in their
models can be traced to the particular form of reference mass 
distribution adopted, which is similar to the relatively peaky 
distribution of the core-forming $n=2$ cloud shown in Figs.~1-3 
(cf. equation [31] of Basu \& Mouschovias 1994). Our more flexible
prescription of mass distribution enables us to uncover a new 
outcome to the ambipolar diffusion-driven evolution of magnetically 
subcritical clouds - ring formation, which has implications on 
cloud fragmentation and multiple star formation. 

\section{DISCUSSION AND IMPLICATIONS} 

\subsection{Ring Fragmentation}
\label{fragmentation}

Dense, self-gravitating rings like the one shown in the right column
of Fig.~3 contain many Jeans masses, and are thus susceptible to 
breaking up gravitationally into a number of smaller pieces. The fact 
that they are gradually condensed out of strongly magnetized clouds 
modifies their fragmentation properties in two important ways. 

First, the strong magnetic fields present during the initial {\it 
magnetically subcritical}, quasi-static phase of cloud evolution 
prevents the Jeans instability from developing on a dynamic time 
scale everywhere, including inside the forming rings (Langer 1978; 
Nakano 1988). On the other hand, after becoming substantially {\it 
supercritical}, the dense rings collapse dynamically, leaving little 
time for small fragments to grow. It therefore appears that the 
best time for the rings to break up gravitationally is during the 
transitional period when they are marginally magnetically critical, 
after being freed from the firm grips of magnetic fields through 
ambipolar diffusion, but before embarking on a runaway collapse. 
If this is indeed the case, then we can estimate the number of 
fragments expected from the breakup. For typical dark cloud conditions, 
the transition occurs roughly around the time $t_1$, when the maximum 
column density reaches 10 times the reference value $\Sigma_0$. 
Since the ring is narrow and self-gravitating at this time, it
should fragment in a way similar to that of an infinitely long
self-gravitating cylinder. It is well known (Larson 1985) that 
(non-magnetic) cylinders with a sound speed $a$ and central mass 
density $\rho_{\rm c}$ are unstable to perturbations with wavelength 
exceeding about
\begin{equation}
\lambda_{\rm cr}=4\left({2 a^2\over \pi G \rho_{\rm c}}\right)^{1/2}.
\label{stab1}
\end{equation}
The instability grows fastest at a wavelength near $2\lambda_{\rm cr}$ 
(Inutsuka \& Miyama 1992). For a ring centered at $r_{\rm r}$, one 
expects the number of fragments to be roughly 
\begin{equation}
N\approx {2\pi r_{\rm r}\over 2\lambda_{\rm cr}}
	\approx {\pi\sigma_{\rm r}\xi_{\rm r}
	\over {16\hat a}^2},
\label{stab2}
\end{equation}
where $\sigma_{\rm r}$ and $\xi_{\rm r}$ are the dimensionless column 
density and radius at the ring location. We have used equations 
(\ref{e7}) and (\ref{stab1}) in deriving 
the second expression, also taking into account the fact that the 
ring is mainly compressed 
by self-gravity. Applying the above formula to the marginally 
critical ring shown in the middle right panel of Fig.~3, we find 
the expected number of fragments to be $\sim 5$ (with 
$\sigma_{\rm r}=10$, $\xi_{\rm r}=0.23$, and ${\hat a}=0.3$). If 
the dimensionless effective sound speed ${\hat a}$ is lowered to 
$0.2$, then the same cloud would produce $\sim 19$ fragments instead
(with $\sigma_{\rm r}=10$ and $\xi_{\rm r}=0.38$). Clearly, the number 
of fragments increases quickly with decreasing sound speed, as one 
might expect intuitively. 

The second modification introduced by a strong magnetic field is 
the possibility of interchange instability. In the simplest case
of an infinitely thin disk supported entirely (and statically) 
by a magnetic field, the square of the growth rate $\gamma$ is 
given by (see Spruit \& Taam 1990)
\begin{equation}
\gamma^2=-{d[\ln(\Sigma/B_z)]\over dr}\cdot g_r, 
\label{e26}
\end{equation}
where $g_r$ denotes the radial component of gravity as before. The 
above equation implies that this (local) instability will grow 
(i.e., $\gamma^2 > 0$) as long as the mass-to-flux ratio, 
$\Sigma/B_z$, decreases in the direction of gravity, $g_r$. This 
criterion will be satisfied in part of the ring during its (long) 
quasi-static phase of formation. The reason is simply that the 
mass-to-flux ratio peaks inside the ring (see Fig.~2a), and thus 
decreases towards the origin, which is also the direction of 
gravity during most of the quasi-static phase of cloud evolution. 
The instability is demonstrated explicitly in Fig.~4, where 
$\gamma^2$ is plotted against the radius and mass for the 
ring-forming $n=8$ cloud shown in Figs.~1-3 at four relatively 
early times: the initial 
equilibrium time $t=0$ and the three times when the maximum column 
density reaches 2, 4, and 8 times the reference value $\Sigma_0$ 
(corresponding to 5.39, 7.76, and 8.52 million years), respectively. 
Note that the inner, plateau part of the cloud (where star formation 
occurs) starts out close to being marginally stable. It becomes 
increasingly more unstable as the ring condenses out, until enough 
mass has accumulated in the ring to reverse the direction of the 
gravity interior to the ring (to outward-pointing). The reversal 
explains the suppression of instability near the center at the last 
time shown. It is likely that magnetic interchange instability, 
which is intrinsic to the ring formation process, contributes 
significantly to ring fragmentation, once the axisymmetry is 
removed. The situation may be complicated, however, by ambipolar 
diffusion and the associated drift between magnetic field lines 
and cloud matter. Non-axisymmetric models are required to examine 
the interplay between magnetic interchange instability and cloud 
fragmentation in detail. 

\subsection{A Scenario of Multiple Star Formation}

Multi-Jeans mass clouds are unstable to forming multiple fragments 
of Jeans mass. The presence of a strong magnetic field in a 
lightly ionized medium such as molecular cloud does not change
the minimum wavelength (and thus mass) for instability, although
the growth time could be greatly lengthened, according to the
linear analysis of Langer (1978). We have shown that nonlinear 
developments of this magnetically-retarded Jeans instability lead,
in a strict axisymmetric geometry, to the formation of dense, 
self-gravitating rings that are prone to breaking up into smaller 
pieces. Generalizing the ring formation and fragmentation 
to non-axisymmetric clouds, we propose the following scenario for 
multiple star formation. As magnetically subcritical, multi-Jeans 
mass clouds evolve due to ambipolar diffusion, dense elongated 
substructures develop quasi-statically. 
These substructures break up, through Jeans and possibly 
magnetic interchange instabilities, into a number of smaller 
pieces, creating a cluster of dense, magnetically supercritical 
cores. The supercritical cores collapse quickly in a (local) 
dynamic time, leading to a burst of star formation.  

The above scenario is a direct extension of the standard picture of 
isolated star formation (Shu et al. 1987; Mouschovias \& Ciolek 
1999) to the formation of multiple stars. A key ingredient is the
gradual condensation of elongated multi-Jeans mass substructures 
capable of breaking up into multiple dense cores. If the breakup 
occurs mostly during the transitional period when the substructure 
is approximately magnetically critical, as we speculated in 
the last subsection, then the initial size of the dense cores and 
the separation between neighboring cores would be comparable to
the Jeans length scale evaluated at the magnetically critical 
density (cf. equation [\ref{stab1}]). The same size scale applies 
to supercritical dense cores formed in isolation, as pointed out 
by Basu \& Mouschovias (1995). 

Although details remain to be worked out, we anticipate two potentially
observable features of the multiple cores formed in the above scenario. 
First, as with single cores formed through ambipolar
diffusion in isolated star formation (e.g., Lizano \& Shu 1989; 
Basu \& Mouschovias 1994), the multiple cores should have magnetic 
field strength close to (say, within a factor of two of) the 
critical value, even after they become supercritical and after 
star formation. Second, the cores should have small motions relative 
to one another, since the motions are cushioned by both the magnetic 
flux trapped inside the cores and the flux held in between them 
(see Fig.~2d). These features are not expected from cores formed, e.g., 
in prompt collapse of non-magnetic Jeans-unstable clouds (Klessen 
et al. 1998). Observations of mass-to-flux ratio and relative motions 
of dense cores can provide key tests of various scenarios of multiple
star formation. 

To make the scenario more quantitative, one needs to construct 
non-axisymmetric models. Such models will allow one to follow 
the ring fragmentation process and, more importantly, to 
study the formation and fragmentation of over dense substructures 
in more realistic clouds that are irregular in shape and/or contain 
appreciable substructures in mass distribution to begin with. 
Moreover, once formed, the fragments (i.e., multiple cores) are 
expected to interact with one another and with their
common envelope, both gravitationally and magnetically. The
interaction should play a role in determining the mass 
spectrum and motions of the cores and thus stars (Motte, Andre
\& Neri 1998; Testi \& Sargent 1998). In particular, 
gravitational drag effect analogous to the ``dynamic friction'' 
in stellar dynamics could in principle bring the cores closer 
together to form binaries and multiple stellar systems 
(Larson 2000). Non-axisymmetric models will also be required 
to follow the nonlinear developments of the (dynamic) Jeans and
magnetic interchange instabilities, which could provide an 
important source of turbulence 
(see also Zweibel 1998). They may explain, at least in part, why 
cluster forming regions are more turbulent than isolated star 
forming regions, even before stars are formed. 

We should stress that the above scenario is intended mainly for the 
formation of stellar groups and small clusters in relatively quiescent 
regions of molecular clouds, 
such as the fragmented starless cores observed in millimeter continuum 
by Ward-Thompson, Motte \& Andre (1999), and perhaps the $\rho$ Oph B2 
core studied in detail by Motte et al. (1998). The $\rho$ Oph B2 
core has 
apparently broken up into a dozen or so starless clumps. It may  
represent a short-lived phase during the evolution of an initially
magnetically subcritical, multi-Jeans mass cloud in which 
supercritical cores have already formed through ambipolar diffusion 
and fragmentation, but yet to collapse into a group of 
stars. The well-studied starless core L1544 may represent a similar
short-lived phase during the formation of an isolated magnetized core 
leading to single star formation (e.g., Williams et al. 1999; Li 1999; 
Ciolek \& Basu 2000). Our scenario for multiple star formation is not 
directly applicable to the formation of rich clusters, such as the 
Trapezium cluster in Orion, where turbulence-induced fragmentation 
probably plays a dominant role (Klessen et al. 1998; Padoan et al. 
2000). 

\subsection{NW Cluster of the Serpens Molecular Cloud Core}

Although one does not expect to find many ring-like structures in
real star forming regions, because of the high degree of geometric 
symmetry required, they are sometimes observed. The most famous 
example is perhaps the ring of massive stars in the massive cluster 
forming region W49N (Welsh et al. 1987). A more recent example is 
the northwestern (NW) cluster of protostars and dense cores in the 
Serpens molecular cloud core -- another active (although less massive) 
cluster-forming region (e.g., Testi \& Sargent 
1998; Davis et al. 1999). The cluster has the appearance of a tilted, 
fragmented ring, and could be resulted from the fragmentation of a 
dense ring-like structure formed quasi-statically in a strongly 
magnetized cloud. This interpretation is strengthened by 
polarization measurements of thermal dust emission, which 
indicates a large scale magnetic field threading the ring plane 
more or less perpendicularly (Davis et al. 2000), as expected 
in our scenario (see Fig.~2d). Moreover, the velocity dispersion 
among the cores is rather small, on the order of 0.25 km s$^{-1}$ 
(J. Williams, personal communication), which could result from 
the magnetic cushion effect mentioned earlier. Infall motions are 
observed on both large, cluster scale ($\sim 0.2$ pc) and small, 
individual core/protostar scale 
($\sim 0.02$ pc; Williams \& Myers 2000). In our picture, the 
large- and small-scale infall motions would be associated, 
respectively,
with the process of ring formation as a whole (similar to that 
shown in Fig.~1b) and with gravitational contraction onto 
individual cores/protostars, although substantial contributions 
from localized turbulence dissipation are also possible (Myers 
\& Lazarian 1998). Our interpretation is complicated, however, 
by the strong turbulent and outflow motions present in the region. 
\vskip 0.75cm

I thank C. Matzner, F. Motte and J. Williams for helpful 
correspondence and an anonymous referee for useful comments. 

\clearpage

\includegraphics[scale=0.75]{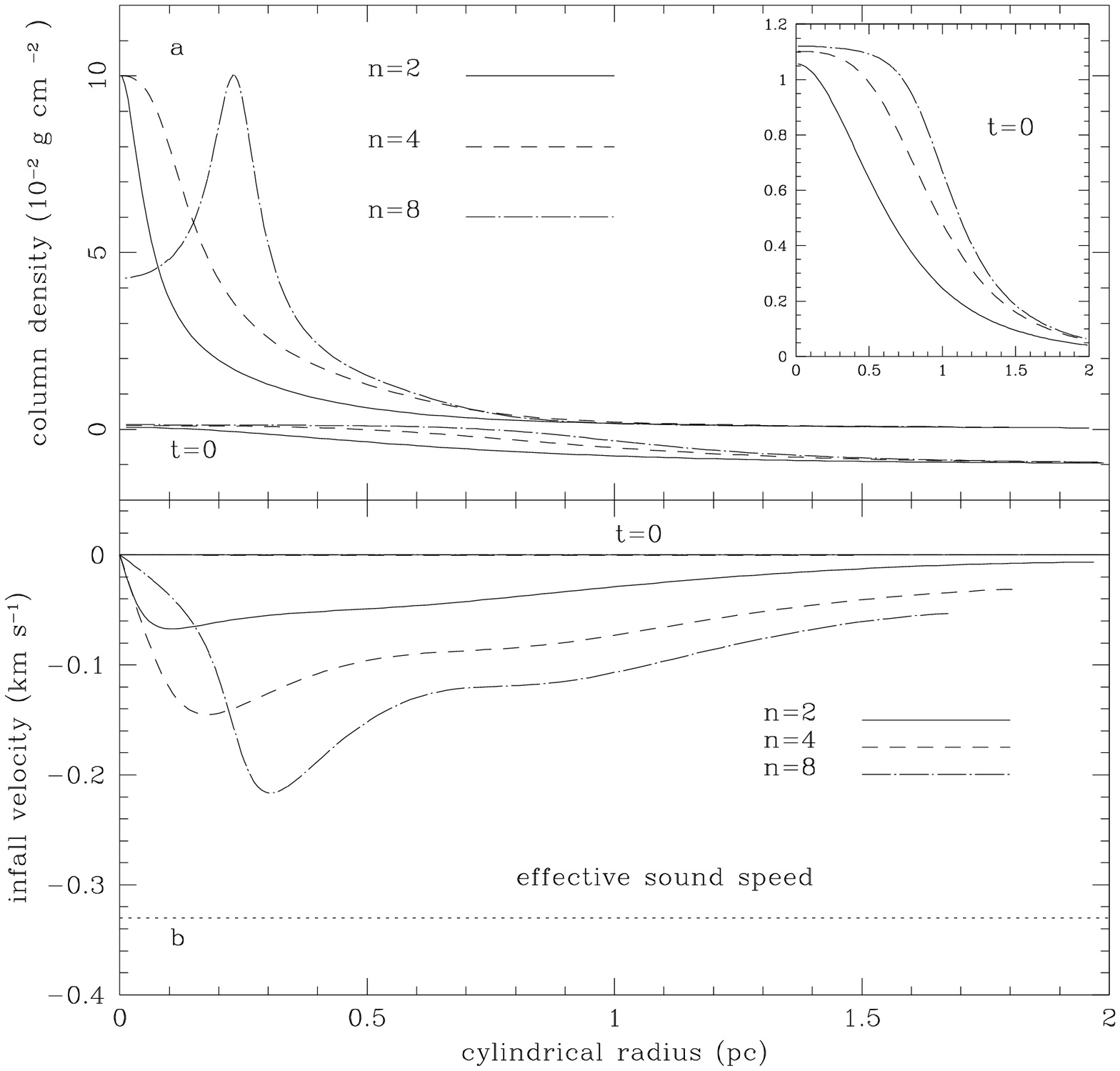}

\begin{figure}

\caption{Evolution of three magnetized clouds with different initial
mass distributions as specified by the exponent $n$ in equation 
(\ref{e9}). (a) Column density distributions at the
initial equilibrium $t=0$ and the time $t_1$ when the maximum column 
density reaches 10 times the reference value $\Sigma_0$. Note that
a dense core is formed at the center of the $n=2$ and $4$ cloud 
whereas a dense ring is produced in the $n=8$ cloud. For clarity,
the distributions at $t=0$ are lowered by one unit. Their differences
show up more clearly in the insert. (b) Distributions of infall 
speed at the same two times as in panel (a), showing the
quasi-static nature of ring and core formation, although substantial
infall motion is clearly present on the (sub-)parsec scale. }
\end{figure}

\clearpage
\includegraphics[scale=0.80]{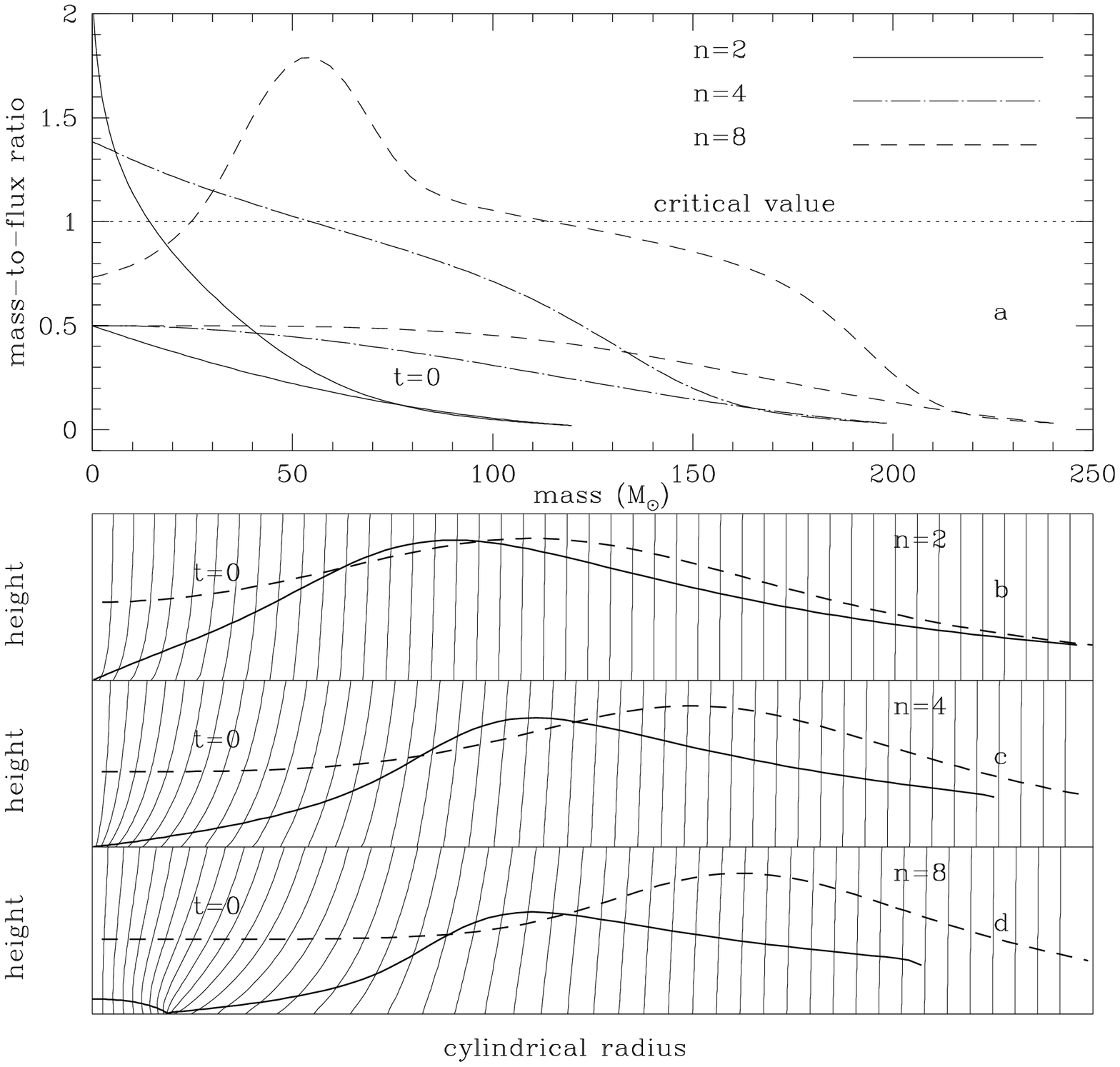}

\begin{figure}

\caption{Additional properties of the model clouds shown in Fig.~1. 
(a) Distributions of 
mass-to-flux ratio, in units of the critical value $(2\pi)^{-1}$
G$^{-1/2}$, as a function of mass at the initial 
time t=0 and the time $t_2$ when the maximum column density reaches 
$10^2$ times the reference value $\Sigma_0$.
Regions above the dotted line are magnetically supercritical.  
(b)-(d) Magnetic field lines at the time $t_2$ (thin solid lines) 
and the disk half thickness at the time t=0 (heavy dashed lines) 
and the time $t_2$ (heavy solid lines). }
\end{figure}

\clearpage

\includegraphics[scale=0.80]{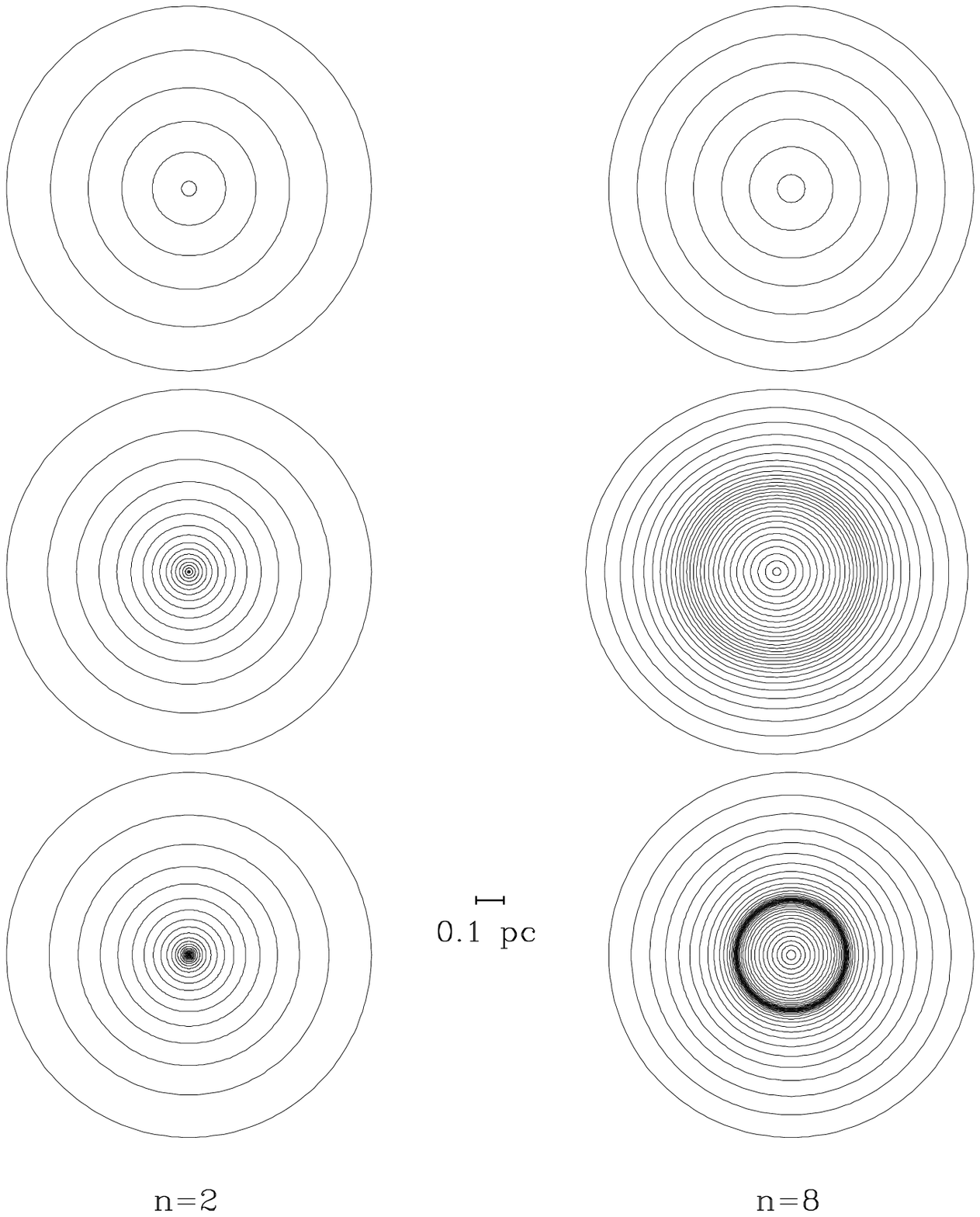}

\begin{figure}

\caption{Graphical representation of core and ring formation in the inner 
region (inside a radius of 0.5 pc) of the $n=2$ (left column) and $n=8$ 
(right column) cloud shown in Figs.~1 and 2. Contours are plotted
at a separation inversely proportional to the local column density. 
They are used to visualize the cloud mass distributions at the
time $t=0$ (top), $t_1$ (middle) and $t_2$ (bottom), when the maximum 
visual extinction through the cloud is $\sim 2$, 20 and 200 magnitudes, 
respectively. 
}
\end{figure}

\clearpage

\includegraphics[scale=0.80]{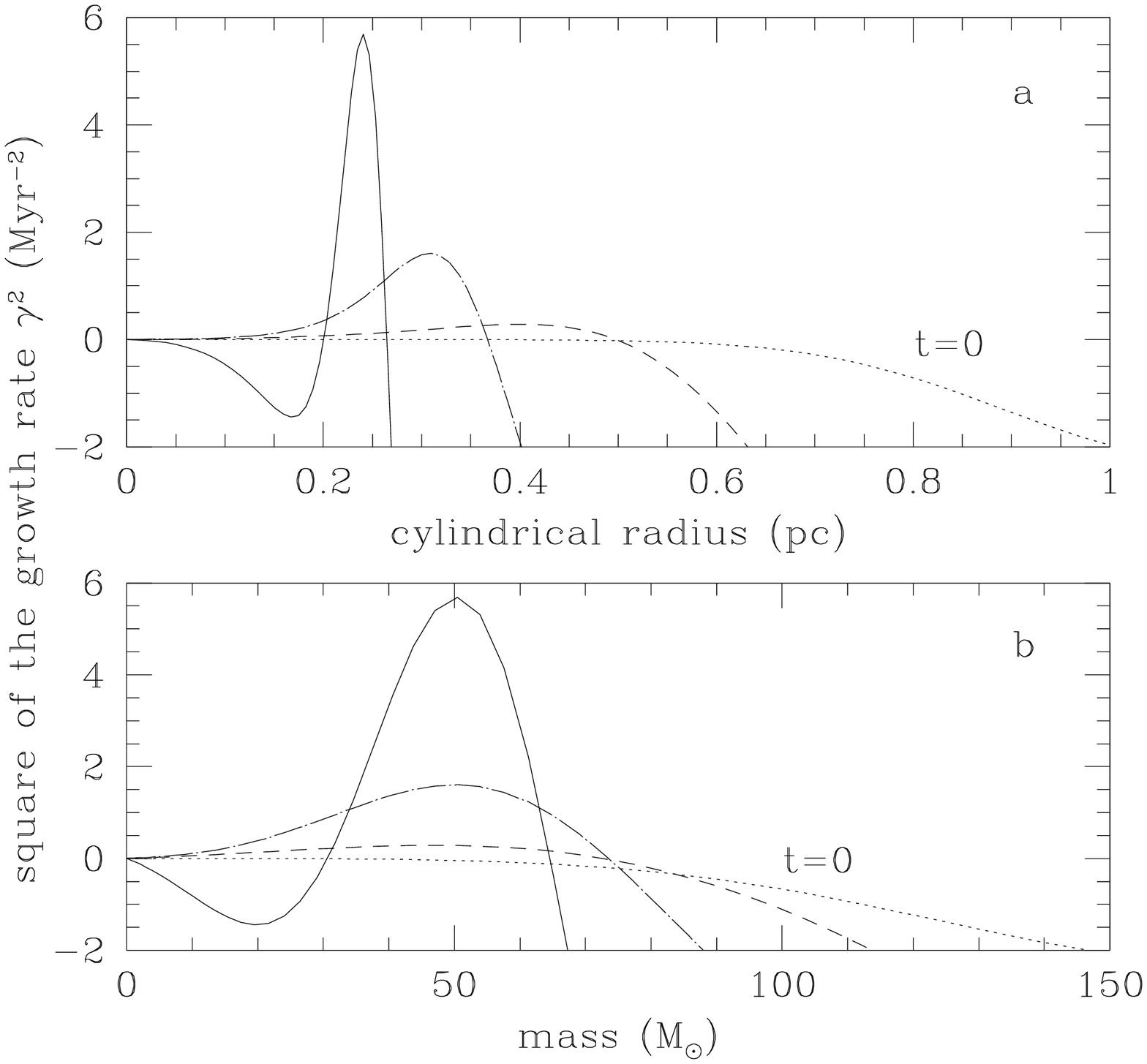}

\begin{figure}
\caption{The square of the growth rate of magnetic interchange 
instability, $\gamma^2$, in units of the inverse of the square 
of a million years, plotted against (a) radius and (b) mass of 
the ring-forming $n=8$ cloud shown in Fig~3, at the initial 
equilibrium time $t=0$ (dotted) and the three times when the 
maximum column density reaches 2 (dashed), 4 (dash-dotted), and 
8 (solid) times the reference value $\Sigma_0$, corresponding 
to 5.39, 7.76, and 8.52 million years. Regions with positive 
$\gamma^2$ are unstable, according to the simple criterion given 
in the text. 
}
\end{figure}

\clearpage
\section{REFERENCES}
\medskip

\noindent
Bastien, P. 1983, AA, 119, 109

\noindent
Basu, S. \& Mouschovias, T. Ch. 1994, ApJ, 432, 720 

\noindent
Beichman, C. A., Myers, P. C., et al. 1986, ApJ, 307, 337

\noindent
Black, D. C. \& Bodenheimer, P. 1976, ApJ, 206, 138

\noindent
Bonnell, I. A. 1999, in The Origins of Stars and Planetary Systems,
eds C. Lada 

\& N. Kylafis (Dordrecht: Kluwer), 479

\noindent
Boss, A. P. 2001, ApJL, in press


\noindent
Bowers, P. J. \& Wilson, J. R. 1991, Numerical Modeling in Applied

Physics and Astrophysics (Boston: Jones and Bartlett Publishers)

\noindent
Ciolek, G. E. \& Basu, S. 2000, ApJ, 529, 925

\noindent
Ciolek, G. E. \& Mouschovias, T. Ch. 1993, ApJ, 418, 774

\noindent
------, 1994, ApJ, 425, 142

\noindent
Crutcher, R. M. 1999, ApJ, 520, 706

\noindent
Davis, C. J., Matthews, H. E., Ray, T. P., Dent, W., Richer, J. S.

1999, MNRAS, 309, 141

\noindent
Davis, C. J., Chrysostomou, A., Matthews, H. E., Jenness, T., Ray,

T. P. 2000, ApJL, in press

\noindent
Fiedler, R. A. \& Mouschovias, T. Ch. 1993, ApJ, 415, 680 

%
\noindent
Inutsuka, S. \& Miyama, S. M. 1992, ApJ, 388, 392

\noindent
Jackson, J. D. 1975, Classical Electrodynamics (New York: 

John Wiley \& Sons)

\noindent
Klessen R. S., Burkert, A., \& Bate, M. R. 1998, ApJ, 501, L205

\noindent
Langer, W. D. 1978, ApJ, 225, 95

\noindent
Larson, R. B., 1985, MNRAS, 214, 379

\noindent
------, 2000, in The Formation of Binary Stars, eds. R. Mathieu \& 
H. Zinnecker, 

in press (astro-ph/0006288)

\noindent
Li, Z.-Y. 1999, ApJ, 526, 806 

\noindent
Lizano, S. \& Shu, F. H. 1989, ApJ, 342, 834

\noindent
Lubow, S. H., Papaloizou, J. C. B., \& Pringle, J. E. 
1993, MNRAS, 267, 235

\noindent
McKee, C. F. 1989, ApJ, 345, 782

\noindent
Motte, F., Andre, P., \& Neri, R. 1998, AA, 336, 150

\noindent
Mouschovias, T. Ch. 1976, ApJ, 206, 753

\noindent
Mouschovias, T. Ch. \& Ciolek, G. E. 1999, in The Origins of Stars and 
Planetary Systems,

eds C. Lada \& N. Kylafis (Dordrecht: Kluwer), p305

\noindent
Mouschovias, T. Ch. \& Morton, S. A. 1991, ApJ, 371, 296

\noindent
Myers, P. C. 1999, in The Origin of Stars and Planet
Systems, eds. C. J. Lada 

\& N. D. Kylafis (Kluwer: Dordrecht), p67

\noindent
Myers, P. C. \& Lazarian, A. 1998, ApJ, 507, L157

\noindent
Nakamura, F., Hanawa, T. \& Nakano, T. 1995, ApJ, 444, 770

\noindent
Nakano, T. 1984, Fundam. Cosmic Phys., 9, 139

\noindent
------, 1988, PASJ, 40, 593

\noindent
------, 1998, ApJ, 494, 587

\noindent
Norman, M. L. \&  Wilson, J. R. 1978, ApJ, 224, 497

\noindent
Norman, M. L. \&  Wilson, J. R., \& Barton, R. T. 1980, ApJ, 239, 968

\noindent
Padoan, P., Nordlund, A., Rognvaldsson, O., Goodman, A. 2000, in From Darkness

to Light, eds. T. Montmerle \& Ph. Andre (ASP Conference Series), in press

\noindent
Pudritz, R. E. 1990, ApJ, 350, 195

\noindent
Shu, F. H. 1992, The Physics of Astrophysics II (Mill Valley:
University Science Books)

\noindent
Shu, F. H., Adams, F. C. \& Lizano, S. 1987, ARAA, 25, 23

\noindent
Shu, F. H., Allen, A., Shang, H., Ostriker, E. \& Li, Z.-Y. 1999, in The 
Origins of Stars and 

Planetary Systems, eds. C. Lada \& N. Kylafis (Dordrecht: Kluwer), p193

\noindent
Shu, F. H. \& Li, Z.-Y. 1997, ApJ, 475, 251

\noindent
Spruit, H. C. \& Taam, R. E. 1990, AA, 229, 475

\noindent
Taffala, M., Mardones, D., Myers, P. C., Caselli, P., Bachiller, R.,

\& Benson, P. J. 1998, ApJ, 504, 900 

\noindent
Testi, L. \& Sargent, A. I. 1998, ApJ, 508, L91

\noindent
Tomisaka, K., Ikeuchi, S., \& Nakamura, T. 1988, ApJ, 335, 239

\noindent
Vazquez-Semadeni, E., Ostriker, E. C., Passot, T., Gammie, C. F., \& Stone, 
J. M. 1999, in 

Protostars and Planets IV, eds. V. Mannings, A. Boss, \& S. Russell
(Univ. Arizona 

Press: Tucson), p3

\noindent
Ward-Thompson, D., Motte, F. \& Andre, P. 1999, MNRAS, 305, 143

\noindent
Welsh, W. J., Jackson, J. M., Dreher, J. W., Terebey, S. \&
Vogel, S. N. 1987, 

Science, 238, 1550

\noindent
Williams, J. P. \& Myers, P. C. 2000, ApJ, in press

\noindent
Williams, J. P., Myers, P. C., Wilner, D. J. \& di Francesco, J. 1999,
ApJ, 613, L61

\noindent
Zweibel, E. G. 1998, ApJ, 499, 746
\end{document}